\documentclass[prb,onecolumn,preprint,superscriptaddress]{revtex4-2}
\usepackage{amsmath,amssymb,bm}
\usepackage{hyperref}
\usepackage{graphicx}
\usepackage{epstopdf}
\usepackage{latexsym}
\usepackage{subfigure}
\usepackage[usenames, dvipsnames]{color}
\usepackage[usenames, dvipsnames]{xcolor}
\usepackage{natbib}
\usepackage{braket}
\usepackage{float}
\usepackage[normalem]{ulem}
\usepackage{comment}
\usepackage{mathtools}
\usepackage{array}
\usepackage{tabu}
\usepackage{multirow}
\usepackage{chemformula}
\usepackage{svg}

\newcommand\redsout{\bgroup\markoverwith{\textcolor{red}{\rule[0.5ex]{2pt}{0.4pt}}}\ULon}
\newcommand\bluesout{\bgroup\markoverwith{\textcolor{blue}{\rule[0.5ex]{2pt}{0.4pt}}}\ULon}

\newcommand{\SPhide}[1]{{}}

\begin{document}
\title{Notes on resummation-based quantum Monte Carlo 
vis-\`a-vis sign-problematic Heisenberg models on 
canonical geometrically frustrated lattices}
\author{Nisheeta Desai}
\affiliation{Department of Theoretical Physics, Tata Institute of Fundamental Research, Mumbai, MH 400005, India}
\author{Sumiran Pujari}
\affiliation{Department of Physics, Indian Institute of
Technology Bombay, Mumbai, MH 400076, India}

\begin{abstract}
We show here that a direct application of resummation-based quantum Monte Carlo (QMC) 
--- implemented
recently for sign-problem-free $SU(2)$-symmetric spin Hamiltonians in the stochastic series expansion 
(SSE) framework ---
does not reduce the sign problem for frustrated $SU(2)$-symmetric
Heisenberg antiferromagnets on canonical geometrically frustrated lattices 
composed of triangular motifs such as the triangular lattice. 
In the process, we demonstrate that resummation-based
updates do provide an ergodic sampling of the SSE-based QMC configurations which
can be an issue when using the standard SSE updates, however, severely
limited by the sign problem as previously mentioned. 
The notions laid out in these notes may be useful
in the design of better algorithms for geometrically frustrated magnets.
\end{abstract}
\maketitle

\newpage 
An outline of these notes: In Sec.~\ref{sec:sse}, we set the stage by briefly
recapitualing the standard SSE set-up and resummation-based updates as they apply
to sign-problem-free spin Hamiltonians on geometrically unfrustrated, bipartite lattices. 
Here, we also give a reason for why one might expect 
resummation to lead to a possible reduction in the sign problem 
for geometrically frustrated, non-bipartite lattices.
In Sec.~\ref{sec:sign_prob_discuss}, we build towards describing 
which kind of SSE-based QMC configurations, 
if they were to exist, would lead to a reduction  of the sign problem with
resummation. We then give an argument for why such configurations do not
exist for Heisenberg antiferromagnets on canonical geometrically frustrated lattices
made out of triangular motifs.
Sec.~\ref{sec:sign_prob_discuss} also
discusses how resummation-based updates allow for an ergodic sampling of the
sign-problematic frustrated QMC ensemble (limited by the sign problem) 
and a closely related sign-problem-free ensemble. 

\section{SSE Basics}
\label{sec:sse}

The SSE framework and SSE-based QMC has been extensively discussed in the literature.
For a detailed exposition, the reader may see the following 
reviews~\cite{Sandvik_2010_review,Sandvik_2019_Ch16review}. Here we limit ourselves
to the basic notions underlying the SSE framework that are required for the 
subsequent sign problem discussions in Sec.~\ref{sec:sign_prob_discuss}.
The equilibrium statistical physics of a quantum Hamiltonian is studied by computing the 
quantum partition function
\begin{equation}
    Z=Tr(e^{-\beta H})
    \label{eq:qpartfunc}
\end{equation}
In SSE, the quantum to classical mapping is carried out by expanding this exponential in a Taylor series. After writing the trace as an expansion in a chosen basis, the partition function is expressed as following:
\begin{equation}
   Z =Tr (e^{-\beta H})=\sum_{\alpha}\langle \alpha | e^{-\beta H}|\alpha \rangle =\sum_{\alpha}\langle \alpha |\sum_n \frac{(-\beta H)^n}{n!} |\alpha \rangle
   \label{eq:sse0}
\end{equation}
where $\{|\alpha\rangle\}$ is the chosen basis. If we introduce a complete set of states between each power of $H$, this series can be written as: 
\begin{equation}
    Z=\sum^{\infty}_{n=0} \frac{\beta^n}{n!} \sum_{\{\alpha\}_n} \langle \alpha_0 |(-H)| \alpha_1 \rangle... \langle \alpha_{n-1}|(-H)|\alpha_0 \rangle
    \label{eq:sse}
\end{equation}

$\{ \alpha \}_n$ indicates there are $n$ states to sum over. 
This brings it into a form where each term in the sum consists of a sequence of spin states : $|\alpha_0\rangle$, $|\alpha_1\rangle$, $|\alpha_2\rangle$...$|\alpha_{n-1}\rangle$, $|\alpha_0\rangle$ with variable $n$. The subscript $i=0,1...,n-1$ in $\alpha_i$ can be related to the ``imaginary time'' dimension and hence a particular sequence of spin states $\{\alpha_i\}$ can be interpreted as a classical spin configuration in one extra dimension.
The different sequences of $\{\alpha_i\}$ are sampled stochastically with their corresponding weight, $W({\{ \alpha \}_n},\beta)$. This weight is proportional to the product of all the matrix elements in Eq.~\ref{eq:sse} evaluated in this sequence as shown below:
\begin{equation}
    W({\{ \alpha \}_n},\beta)=\frac{\beta^n}{n!} \prod^{n}_{i=0} \langle \alpha_i |(-H)| \alpha_{i+1} \rangle
    \label{eq:weightsse}
\end{equation}
where $|\alpha_{n+1}\rangle=|\alpha_0\rangle$. For an arbitrary $H$, however, every term in the series in Eq.~\ref{eq:sse} need not have a positive weight resulting in the notorious sign problem. 
For a Hamiltonian to be amenable to QMC simulations, a sufficient condition is
the following: 
\begin{equation}
 \langle \alpha_{i} | H | \alpha_{j} \rangle \leq 0 \;\;\;\;\; \forall \; i,j 
 \label{eq:signfree}
\end{equation}
The diagonal matrix elements can be made negative by subtracting a large constant, therefore the above condition really concerns the off-diagonal matrix elements when designing or discussing algorithms. This condition is equivalent to the Marshall sign rule.
The Heisenberg antiferromagnet on a bipartite lattice, is an example of a sign-problem-free model, where these classical configurations can be further interpreted as 
closely-packed loop configurations.

\subsection{The Heisenberg antiferromagnet as an example} \label{subsec:intro-heisantiferro}

\begin{figure}[t]
 \centering
 \includegraphics[width=0.6\linewidth]{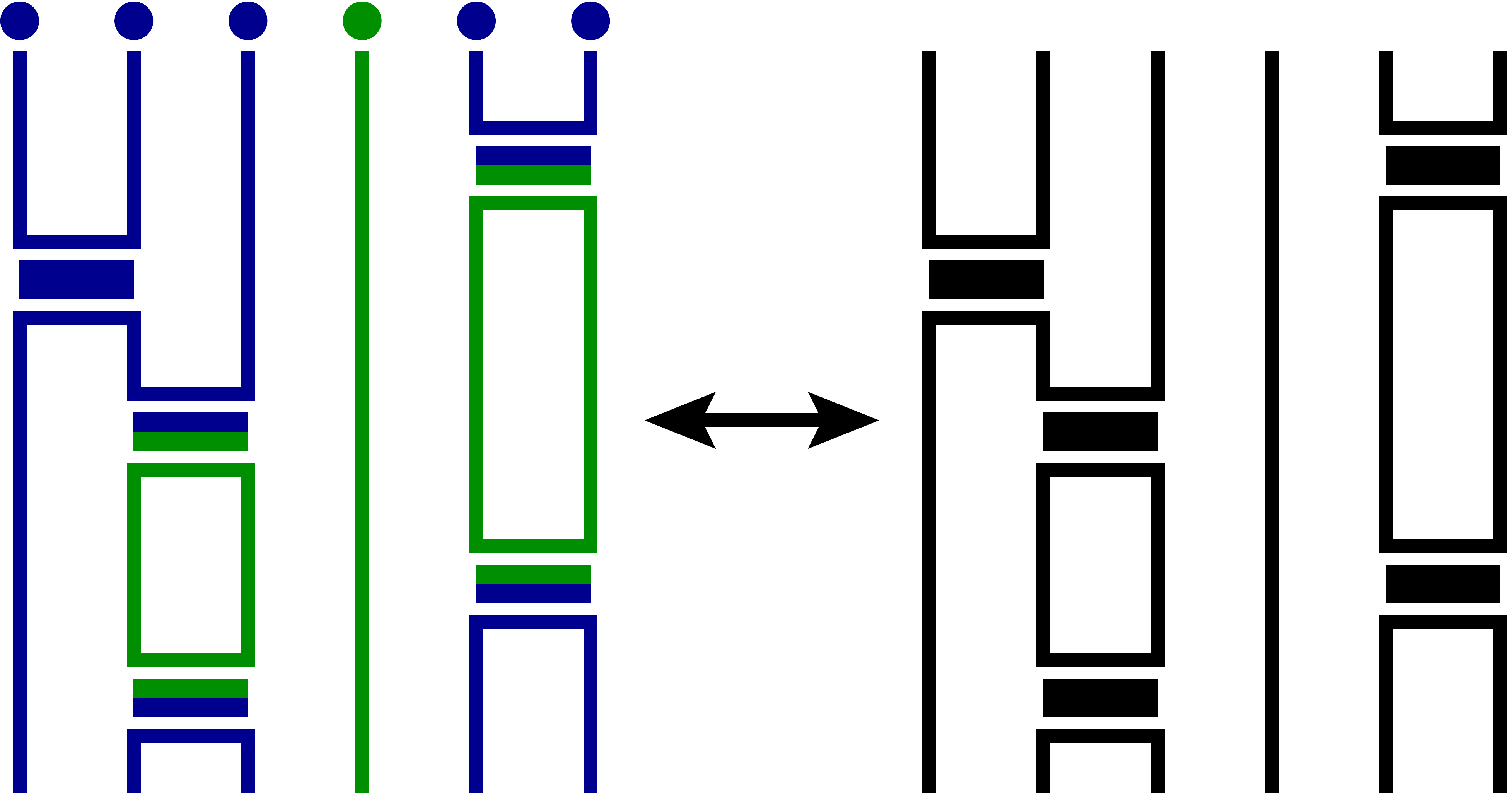}
 \caption{\label{fig:loop_diag} An example of a QMC configuration with closely-packed loops in the
SSE framework.}
\end{figure}

The Heisenberg interaction between two sites, upto a constant $S^2$, can be written as : 
\begin{equation}
    H^{ij}_2= - J (S^2-\vec{S_i}.\vec{S_j})
    \label{eq:heisint}
\end{equation}{}
where $S$ is the value of the spin. The subtraction of $S^2$ makes all diagonal elements negative and a unitary transformation on one of the sites, $S_j^x \rightarrow -S_j^x$ and $S_j^y \rightarrow -S_j^y$, makes all the off-diagonal elements of $H^{ij}_2$ negative satisfying the condition of Eq.~\ref{eq:signfree}. By carrying out this unitary transformation on one of the sublattices of the bipartition, the Heisenberg interaction on a bipartite lattice can be made sign-problem-free. 
On a non-bipartite lattice, if this transformation is carried out given some bipartition, the model still retains a sign problem due to $H^{ij}_2$ on bonds connecting sites in the same sublattice.

For $S=\frac{1}{2}$, the Heisenberg interaction after the above unitary transformation can be written as:
\begin{align}
 H^{ij}_2 = -\frac{J}{2} \left( |\uparrow^z_i \downarrow^z_j\rangle \langle \uparrow^z_i \downarrow^z_j| + |\downarrow^z_i \uparrow^z_j \rangle \langle \downarrow^z_i \uparrow^z_j| + |\uparrow^z_i \downarrow^z_j\rangle \langle \downarrow^z_i \uparrow^z_j|+ |\downarrow^z_i \uparrow^z_j \rangle \langle  \uparrow^z_i \downarrow^z_j| \right)
 \label{eq:heisspinhalf}
\end{align}
with $\{ \uparrow^z_i, \downarrow^z_i \}$ denoting the $S^z$ eigenstates at site $i$.
The nearest neighbour Heisenberg Hamiltonian can now be written as a sum on bond operators:
\begin{equation}
    H=\sum_{\langle ij \rangle}H^{ij}_2 \equiv \sum_{b} H_b
    \label{eq:sumonbonds}
\end{equation}
The powers of $H$ in Eq.~\ref{eq:sse0} can then be replaced by a sum on product strings of two body operators, $H_{b_1}H_{b_2}....H_{b_n}$. If $S_n$ denotes a string of operator indices, the partition function becomes:
\begin{equation}
    Z=\sum^{\infty}_{n=0} \frac{(-\beta)^n}{n!} \sum_{S_n} \sum_{\alpha} \langle \alpha | H_{b_1}H_{b_2}\hdots H_{b_n}| \alpha \rangle
    \label{eq:sseheis}
\end{equation}
The ``computational" basis states, $\alpha$, usually are chosen as the tensor product states: 
\begin{equation}
    |\alpha\rangle=| \mathcal{S}_0 \rangle \otimes | \mathcal{S}_1 \rangle \otimes | \mathcal{S}_2 \rangle \otimes \hdots 
    \label{eq:alpha}
\end{equation}
with $\mathcal{S}_i \in \{ \uparrow^z_i, \downarrow^z_i \}$. 
The bond operators, $H_b$, are chosen such that the action of the operator on one basis state gives another basis state and not a linear combination of them, i.e. $H_b|\alpha\rangle=|\alpha'\rangle$, with $|\alpha\rangle$ and $|\alpha'\rangle$ both being $S^z$ basis states. This is done by separating the diagonal and off-diagonal operator pieces of $H^{ij}_2$, i.e. the index $b$ does not just refer to the bond location, but also for distinguishing a diagonal operator action from an off-diagonal operator action in Eq.~\ref{eq:heisspinhalf}. This so-called ``no branching'' condition~\cite{Sandvik_2019_Ch16review} ensures that the successive action of operators in the operator string on $|\alpha\rangle$, yields a unique sequence of states in ``imaginary time''. Thus, the sum on all sequences of spin states in Eq.~\ref{eq:sse} is replaced by an equivalent sum on operator strings. Sampling spin configurations corresponding to Eq.~\ref{eq:sseheis} can be shown to be equivalent to sampling closely-packed coloured loop configurations in one higher dimension~\cite{Sandvik_2019_Ch16review}. For certain models, like rotationally symmetric $S=\frac{1}{2}$ models, every operator string corresponds to a unique loop configuration and vice versa as shown in the left panel of Fig.~\ref{fig:loop_diag}. Each of the loops in the closely-packed coloured loop configuration can be coloured independently. The number of colours the loops can have is equal to the number of ways one can ``colour'' a spin or the basic degree of freedom that lives on the lattice sites. In this case the spins can have two states, $|\uparrow^z\rangle$ or $|\downarrow^z \rangle$, therefore the loops can have two colours.
We may then rewrite the partition function as 
\begin{equation}
 Z= \sum_{\{ C_{\text{loops}}\}} \sum_{\{\alpha\}} W(C_{\text{loops}})
 \label{eq:clpartitionfunc}
\end{equation}
Here $C_{\text{loops}}$ is any allowed closely-packed uncoloured loop-gas configuration with only one underlying operator where two loops abut each other  at various
(imaginary) time slices (Fig.~\ref{fig:loop_diag} right). The weight of each such configuration is given by, $W(C_{\text{loops}})$, where $W(C_{\text{loops}}) = \frac{1}{n!} \left( \frac{\beta J}{2}\right)^n$. $\{\alpha\}$ is a set of colours for the loops. The left panel of Fig.~\ref{fig:loop_diag} shows one example from this set. If there are $n_l$ loops in the configuration, the sum over $\{\alpha\}$ is over $2^{n_l}$ coloured configurations corresponding to the independent ways of colouring all the loops. The SSE algorithm involves sampling the coloured loop configurations according to their weights. Changing the colour of a loop leads to a non-local update (sometime called an ``operator loop" update or just a loop update) in the QMC configuration which makes the SSE approach quite powerful.

Now, we may go one step further and resum over the spin or colour values of these closely-packed loops without breaking or changing any loop connections 
in the operator string. This 
then renders the ensemble as a configuration of closely-packed uncoloured
loops as shown in the right panel of Fig.~\ref{fig:loop_diag}. This 
loop-gas representation for the high-temperature series may be written as
\begin{equation}
    Z(\beta)=\sum^{\infty}_{n=0} \frac{(-\beta)^n}{n!} 2^{n_l}
    \sum_{S_n} h_{b_1} h_{b_2} \hdots h_{b_n}
    \label{eq:loopgas_rep}
\end{equation}
where the bond index $b_i$ is now the only indexing required, $h_{b_i}$ indicates the spin-symmetric matrix element contribution ($-\frac{J}{2}$ in our example) at $b_i$, and $n_l$ is the number of loops in a given configuration.
Eq~\ref{eq:clpartitionfunc} is then modified to 
\begin{equation}
 Z= \sum_{\{ C_{\text{loops}}\}} W(C_{\text{loops}})
 \label{eq:clpartitionfunc}
\end{equation}
where $C_{\text{loops}}$ is as defined earlier, but the weights are modified to $W(C_{\text{loops}}) = \frac{2^{n_l}}{n!} \left( \frac{\beta J}{2}\right)^n $.

The algorithm to sample uncoloured loop configurations was dubbed as the resummed-SSE (RSSE) algorithm~\cite{Desai_Pujari_prb2021}. 
This has some advantages over the standard or coloured SSE algorithm. It is 
naturally advantageous in simulating quantum paramagnetic phases where coloured SSE can face ergodicity problems. It has been shown in previous work that partial resummation can also be used as a solution to the sign problem for certain models in world-line loop algorithms~\cite{Chandrasekharan_Wiese_prl1999,Henelius_Sandvik_prb2000} which is a closely-related algorithm to SSE~\cite{Sandvik_2010_review}. In these models, when one flips the colour of certain coloured loops, it changes the overall sign of the QMC configuration without changing its weight. Such loops were called ``merons''. Resumming over merons can eliminate the sign problem in certain models. In Sec.~\ref{sec:meron_resum}, we describe one quantum spin model (perhaps the only) where this technique has been shown to be successful in eliminating the sign problem for certain observables and minimizing it in others. This technique cannot be blindly applied to models on geometrically frustrated lattices because these algorithms are often not ergodic on such lattices as we will see through examples. In Sec.~\ref{sec:ergodicity} we describe how this problem can be remedied using the RSSE algorithm. It can thus be used to simulate the Heisenberg model on canonical geometrically frustrated lattices ergodically. However, this still has a severe sign problem.  In Sec.~\ref{sec:sign_prob}, we explore a possible approach to reduce the sign problem on these lattices based on meron resummation and explain why it does not work. 

\section{Sign-problem discussion}
\label{sec:sign_prob_discuss}

\subsection{Meron resummation}
\label{sec:meron_resum}
Consider the following $XXZ$ Hamiltonian, with a ferromagnetic coupling in the $z$-direction and an antiferromagnetic coupling in the $xy$-plane:
\begin{equation}
 H=-\sum_{\langle ij \rangle} (S^z_i S^z_j- S^x_i S^x_j - S^y_i S^y_j)
 \label{eq:semifrus_model}
\end{equation}
This model has positive off-diagonal matrix elements. This is equivalent to the Heisenberg ferromagnet on a bipartite lattice upto the unitary transformation on one sublattice of the bipartition. On a frustrated lattice, this model has a severe sign problem. One 
way to remedy the sign problem is to change the basis~\cite{Henelius_Sandvik_prb2000}. By relabeling such that the ferromagnetic coupling is in the $y$-direction, we get:
\begin{equation}
 H=-\sum_{\langle ij \rangle} (S^y_i S^y_j- S^x_i S^x_j - S^z_i S^z_j)
\end{equation}
Now working in the basis where $S^x$ is diagonal, we define $S^+=S^y+iS^z$ and $S^-=S^y-iS^z$. $H$ can be written, upto a constant, as:  
\begin{equation}
 H=-\sum_{\langle ij \rangle} \Bigg[ \Bigg(\frac{1}{4}- S^x_i S^x_j\Bigg) + \frac{1}{2} (S^+_i S^+_j + S^-_i S^-_j) \Bigg]
\end{equation}
This can be rewritten in terms of $S^x$ basis states:
\begin{figure}[t]
 \includegraphics[width=0.5\linewidth]{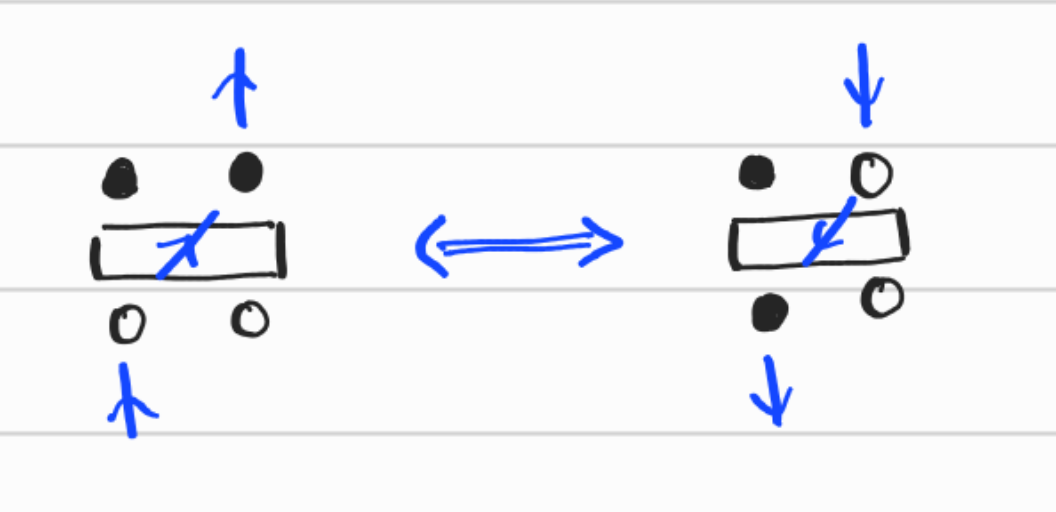}
 \caption{\label{fig:semifrus_xbasis} Loop moves in the operator loop update for the model in Eq.~\ref{eq:semifrus_xbasis}. }
\end{figure}
\begin{align}
 H = -\frac{1}{2} \sum_{\langle ij \rangle} (|\uparrow^x_i\downarrow^x_j\rangle\langle\uparrow^x_i\downarrow^x_j|+|\downarrow^x_i\uparrow^x_j\rangle\langle\downarrow^x_i\uparrow^x_j|+
 |\uparrow^x_i\uparrow^x_j\rangle\langle\downarrow^x_i\downarrow^x_j| +
 |\downarrow^x_i\downarrow^x_j\rangle\langle\uparrow^x_i\uparrow^x_j|)
 \label{eq:semifrus_xbasis}
\end{align}
All matrix elements are negative, hence this is now a sign-problem-free model. The loop moves in the operator loop update for this model are shown in Fig~\ref{fig:semifrus_xbasis}.

\begin{figure}[h]
 \includegraphics[width=0.5\linewidth]{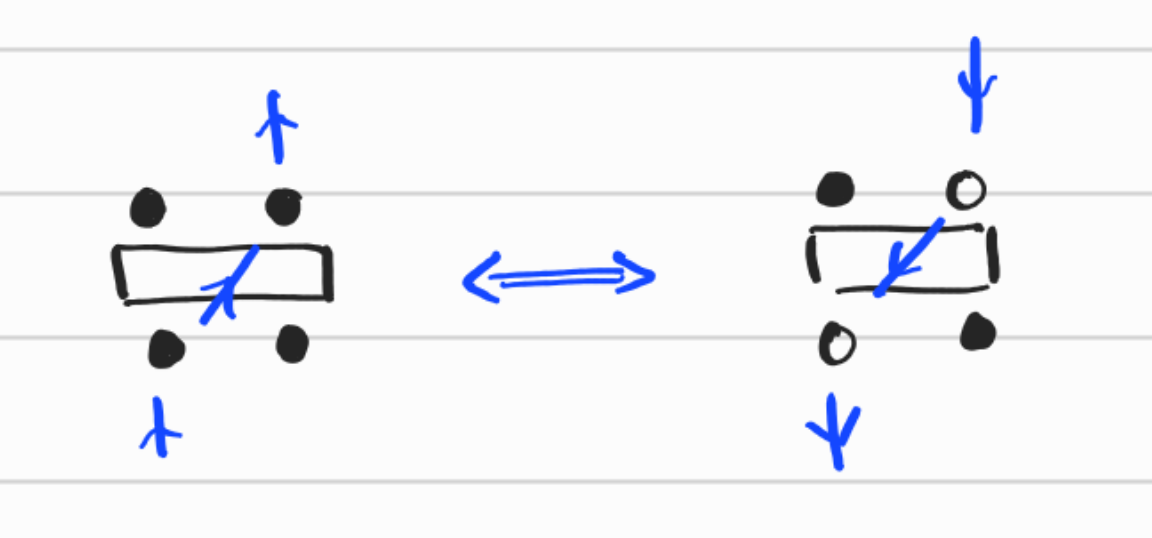}
 \caption{\label{fig:semifrus_zbasis} Loop moves in the operator loop update for the model in Eq.~\ref{eq:semifrus_zbasis}. }
\end{figure}

Another solution to the sign problem using the concept of a ``meron'' or sign-changing loop was given by Henelius and Sandvik on the square lattice with nearest and next-nearest neighbour interactions~\cite{Henelius_Sandvik_prb2000} which we briefly describe now. Using this concept, the model can actually be simulated in the original basis. In terms of $S^z$ basis states, $H$ can be written, upto a constant, as:
\begin{align}
 H=-\frac{1}{2}\sum_{\langle ij \rangle} (|\uparrow^z_i\uparrow^z_j\rangle\langle\uparrow^z_i\uparrow^z_j|+|\downarrow^z_i\downarrow^z_j\rangle\langle\downarrow^z_i\downarrow^z_j|-
 |\uparrow^z_i\downarrow^z_j\rangle\langle\downarrow^z_i\uparrow^z_j|-|\downarrow^z_i\uparrow^z_j\rangle\langle\uparrow^z_i\downarrow^z_j|)
 \label{eq:semifrus_zbasis}
\end{align}
Loop moves in the operator loop update are shown in Fig~\ref{fig:semifrus_zbasis}. The net contribution from off-diagonal operators is sign-problematic when simulated on a frustrated lattice. (On a bipartite lattice, the net contribution is sign-problem-free by virtue of loops necessarily involving even number of off-diagonal operators.) Flipping the colour of any loop converts the off-diagonal operators touched by the loop into diagonal operators and vice versa. Therefore, upon flipping the colour of loops touching an odd number of operators leads to a change in the sign of the overall QMC configuration. Therefore, such loops serve as ``merons'' and the weights of configurations with non-zero merons will necessarily resum to zero under colour flips. Sampling over the zero meron sector was shown to eliminate the sign problem for certain important observables~\cite{Henelius_Sandvik_prb2000}.

\subsection{Ergodicity problem on geometrically frustrated lattices and RSSE}
\label{sec:ergodicity}

Consider the following two-spin $XXZ$ interaction on a frustrated lattice built out of triangular motifs:
\begin{equation}
 H^{ij}_{XXZ} = S^z_i S^z_j - S^x_i S^x_j - S^y_i S^y_j   
\end{equation}
This is the negative of the interaction in Eq~\ref{eq:semifrus_model}. This model has all negative off-diagonal matrix elements, and therefore it is sign-problem-free. It can be written as:
\begin{align}
 H^{ij}_{XXZ} = - \frac{1}{2} (|\uparrow^z_i \downarrow^z_j \rangle \langle \uparrow^z_i \downarrow^z_j| + |\downarrow^z_i \uparrow^z_j \rangle \langle \downarrow^z_i \uparrow^z_j|  
 +|\uparrow^z_i \downarrow^z_j \rangle \langle \downarrow^z_i \uparrow^z_j | + |\downarrow^z_i \uparrow^z_j \rangle \langle \uparrow^z_i \downarrow^z_j | )
\end{align}
Naively, this can be simulated without a sign problem on a non-bipartite lattice. 
However, the standard SSE algorithm is not ergodic in simulating this model on a geometrically frustrated lattice. This can be seen by simulating the following model on a frustrated chain:
\begin{equation}
 H = \sum_{\langle ij \rangle} H^{ij}_{XXZ} + g \sum_{\langle \langle ij \rangle \rangle}  H^{ij}_{XXZ}
 \label{eq:triangchain2}
\end{equation}
\begin{table}
\begin{tabular}{|c|c|c|c|c|c|}
 \hline
 g & $\beta$ & $e_1$ (ED) & $e_1$ (SSE) & $e_1$ (RSSE) & $e_2$ (SSE) \\
 \hline
  0.1 & 1.0 & -0.485421 & -0.4774(2) & -0.4854(4)  & -0.4858(6) \\
  \hline
  0.5 & 1.0 & -0.647273 & -0.6105(2) & -0.6473(4) & -0.6476(6)  \\
  \hline
  1.0 & 1.0 & -0.938847 & -0.8769(4) & -0.9387(4)  & -0.9378(7)  \\
  \hline
\end{tabular}
\caption{\label{tab:energy_EDvsQMC}Comparison of energy per unit site, $e_1$, of model in Eq.~\ref{eq:triangchain2} for $L=8$ calculated from exact diagonalization (ED), SSE and RSSE. The SSE values clearly do not match ED but RSSE values do. $e_2$ denotes energy per unit site of the model in Eq.~\ref{eq:triangchain2} where $H^{ij}_{XXZ}$ is replaced by Eq.~\ref{eq:XXZ2}, using SSE. Comparison with ED shows that this model gives correct results. }
\end{table}
By comparing the energy observable per unit site for different parameters as given in Table~\ref{tab:energy_EDvsQMC}, the lack of ergodicity is seen clearly. Compare the third and fourth columns of Table~\ref{tab:energy_EDvsQMC}.
\begin{figure}[b]
 \includegraphics[width=0.3\linewidth]{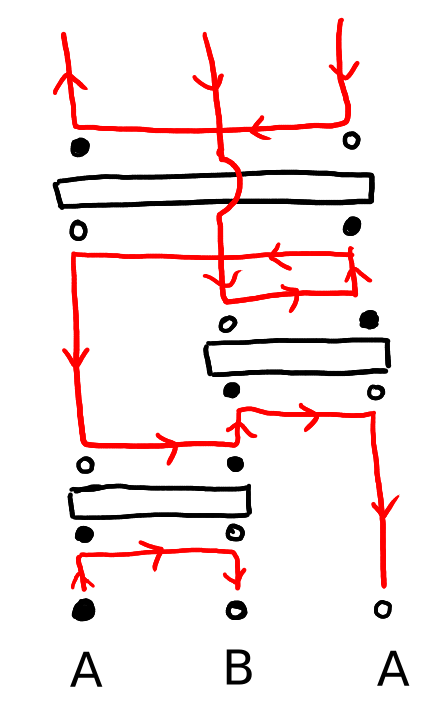}
 \caption{\label{fig:non_ergodic_eg} An example of a QMC configuration in space-time having an odd number of off-diagonal operators.}
\end{figure}
The standard SSE loop update is unable to generate certain configurations in the configuration space of this model. These are configurations with an odd number of off-diagonal operators, an example of such a configuration is shown in Fig.~\ref{fig:non_ergodic_eg}. The SSE updates principally involve two steps: (1) a diagonal update, involving insertion and removal of diagonal operators, and (2) the operator loop update that converts diagonal operators into off-diagonal operators and vice versa. In this model, any loop has to touch an even number of operators in order for it to close. Since, an even number is either the sum of two even numbers or two odd numbers, therefore flipping a loop can only make an even number of diagonal operators into off-diagonal operators (and vice versa) or an odd number of diagonal operators into off-diagonal operators (and vice versa). In other words, the loop update cannot change the parity of the count of off-diagonal operators in any configuration. One generally starts with a configuration consisting of all diagonal operators, from which it is impossible to generate a configuration consisting of an odd number of off-diagonal operators with standard SSE updates. This is how ergodicity gets violated by being stuck in one of these parity sectors. 
This problem can be circumvented by a change of basis similar to the one in the previous section:
\begin{align}
 H^{ij}_{XXZ}= -\Big(\frac{1}{4}+S^x_i S^x_j\Big)-(S^y_iS^y_j-S^z_iS^z_j) 
             = -\Big(\frac{1}{4}+S^x_i S^x_j\Big)-\frac{1}{2}(S^+_iS^+_j-S^-_iS^-_j)
 \label{eq:xxzham}
\end{align}
In terms of the $S^x$ basis states, this can be written as:
\begin{align}
 H^{ij}_{XXZ}=-\frac{1}{2} \left( |\uparrow^x_i\uparrow^x_j\rangle\langle \uparrow^x_i \uparrow^x_j|+|\downarrow^x_i\downarrow^x_j\rangle\langle \downarrow^x_i \downarrow^x_j|  + |\uparrow^x_i\uparrow^x_j\rangle\langle \downarrow^x_i \downarrow^x_j|+|\downarrow^x_i\downarrow^x_j\rangle\langle \uparrow^x_i \uparrow^x_j| \right)
\label{eq:XXZ2}
\end{align}
This model is fully ergodic on frustrated lattices and does not suffer from a sign problem. This is because for this model there can not be configurations with an odd number of off-diagonal operators similar to the example in Fig.~\ref{fig:non_ergodic_eg}. In fact, the example in Fig.~\ref{fig:non_ergodic_eg} is the geometric space-time primitive or base case required to have an odd number of off-diagonal operator in any configuration for models with two-site Heisenberg terms in the Hamiltonian. Any model which is not compatible with such a space-time primitive will not have an odd number of off-diagonal operators. Thus, only the even parity sector is involved here and the loop update can sample it ergodically. See the sixth column of Table~\ref{tab:energy_EDvsQMC}.

Another solution to simulate the model in the $S^z$ basis itself is to use RSSE which overcomes the ergodicity limitation of SSE. In RSSE, there is no notion of diagonal and off-diagonal operators as we saw in Sec.~\ref{sec:sse}, which effectively makes this algorithm ergodic in sampling both parity sectors for the off-diagonal operator counts. If one carries out the update in a colour-blind way and then colours the loops, one can easily generate configurations as shown in Fig~\ref{fig:non_ergodic_eg} and others with an odd number of off-diagonal operators. See the fifth column of Table~\ref{tab:energy_EDvsQMC}. 

\subsection{Resummation vis-\`a-vis the sign problem due to geometric frustration}
\label{sec:sign_prob}

Consider finally the following Hamitonian: 
\begin{equation}
 H = \sum_{\langle ij \rangle} H^{ij}_2 + g \sum_{\langle \langle ij \rangle \rangle}  H^{ij}_2
 \label{eq:triangchain}
\end{equation}
involving just Heisenberg terms. Here, the interaction of Eq~\ref{eq:heisint} is summed over nearest-neighbour bonds and the next-nearest-neighbour bonds of a 1$d$ chain. This chain model is thus a geometrically frustrated system. If  we label the even and odd sites on the chain as A and B, the unitary transformation on A sites (say), $S^x_j \rightarrow -S^x_j$ and $S^y_j \rightarrow -S^y_j$, does not make the model sign-problem-free. For an AB bond, the interaction can be written in the $S^z$ basis as in Eq~\ref{eq:heisspinhalf} which is sign-problem-free. For an AA bond, the off-diagonal terms change sign:
\begin{align}
 H^{ij}_{2,\text{AA}} = -\frac{J}{2} (|\uparrow^z_i \downarrow^z_j\rangle \langle |\uparrow^z_i \downarrow^z_j| + |\downarrow^z_i \uparrow^z_j \rangle \langle \downarrow^z_i \uparrow^z_j| - |\uparrow^z_i \downarrow^z_j\rangle \langle |\downarrow^z_i \uparrow^z_j| - |\downarrow^z_i \uparrow^z_j \rangle \langle  \uparrow^z_i \downarrow^z_j| )
\end{align}
Thus, the sign problem still remains due to off-diagonal operators on AA bonds coming from QMC configurations with an odd number of such AA off-diagonal operator contributions as in the example of Fig.~\ref{fig:non_ergodic_eg}. In a similar vein to Sec.~\ref{sec:ergodicity}, RSSE can sample this QMC ensemble ergodically as shown in Table~\ref{tab:energy_EDvsQMC}, while standard SSE updates do not ergodically sample both parity sectors for the off-diagonal operator counts.
\begin{table}[t]
\begin{tabular}{|c|c|c|c|}
 \hline
 g & $\beta$ & $e$ (ED) & $e$ (RSSE) \\
 \hline
  0.1 & 0.01 & -0.276897 & -0.27691(3)  \\
  \hline
  0.1 & 0.025 & -0.279755 & -0.2797(1) \\
  \hline
  0.1 & 0.05 & -0.284548 & -0.283(2) \\
  \hline
  0.2 & 0.01 & -0.301952 & -0.30193(4) \\
  \hline
  0.2 & 0.025 & -0.304887  & -0.3048(3) \\
  \hline
  0.2 & 0.05 & -0.309794  & 0.3(1) \\
  \hline
\end{tabular}
\caption{\label{tab:energy_EDvsQMC} Comparison of energy per unit site, $e$, of model in Eq.~\ref{eq:triangchain} for $L=8$ calculated from ED and RSSE. The statisical estimation error clearly shoots up rapidly as the temperature is reduced or $g$ is increased.}
\end{table}

Let us hypothetically have a loop in the QMC configuration that touches an odd number of AA bond operators (if the loop touches the same AA bond operator twice, it is counted as 2).
When the colour of such a loop is flipped, the off-diagonal operators become diagonal and vice versa. Since 1) an odd number is always a sum of an even and an odd number, and 2) the operators on the AA bonds change sign when undergoing a change from diagonal action to off-diagonal action or vice versa, therefore, the weight of the full QMC configuration changes sign under such a colour flip. As discussed previously, such loops that change sign upon colour flips can serve as merons~\cite{Henelius_Sandvik_prb2000} now in this context. If such loops were to exist, they can be resummed away to reduce the sign problem in simulations of this model, as has been described before for a semi-frustrated model in Sec.~\ref{sec:meron_resum}.

\begin{figure}[t]
 \includegraphics[width=0.3\linewidth]{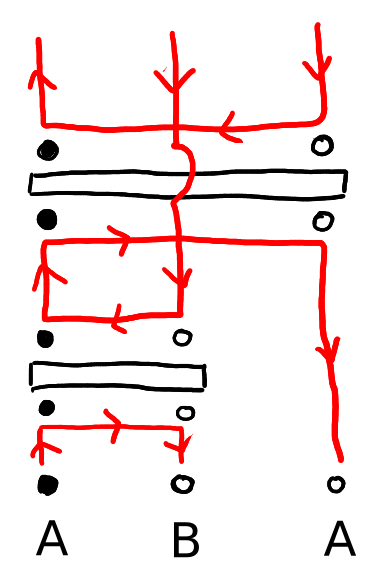}
 \caption{\label{fig:AA_bond_eg} Orientation of loop as it encounters bond operators. The loops always switch direction on encountering a bond. The nearest-neighbour bonds are AB bonds and next-nearest-neighbour bonds are AA bonds. The loop is required to touch an even number of AA bonds so as to come back to the same direction as it started.}
\end{figure}

Such loops, however, don't exist in the QMC configuration space of Eq.~\ref{eq:triangchain} as can be explicitly checked in simulations. 
This is also the case for other geometrically frustrated lattices like the triangular lattice, the kagome lattice, etc.
This can be understood by providing the loops with an orientation. When one starts to grow a loop, the direction of the loop switches when an operator is encountered (see Figs~\ref{fig:AA_bond_eg} and~\ref{fig:non_ergodic_eg}). Encountering an AB bond also switches the sublattice along with the direction, therefore it is clear that the loop has to encounter an even number of AB bond operators
to come back to the same sublattice it started with in order to close the loop. If the model had only AB bonds (as in the bipartite case), the loop will be oriented in one direction on the A sublattice and in the opposite direction on the B sublattice. Therefore, when the loop comes back to the point it started, it is automatically ensured that it does with the same orientation as the one it started with, as the orientation is determined by the sublattice of the starting point. Note that loops do not retrace themselves in this model and many-related antiferromagnetic models. 
In presence of AA bond operators, however, the orientation is not determined by the sublattice. These bonds reverse the direction of the loop on the same sublattice (see Fig~\ref{fig:AA_bond_eg}). Therefore, in order to ensure that the direction of the loop is reversed twice on the same sublattice to close the loop, it is required to touch an even number of AA bond operators. This forbids the hypothesized merons thus precluding a meron resummation strategy to reduce the sign problem in canonical geometrically frustrated lattices.

\begin{acknowledgements}
N.D. was supported by a National Postdoctoral Fellowship from
SERB-DST, Govt. of India (PDF/2020/001658)
hosted at the Dept of Theoretical Physics of the Tata Institute of Fundamental Research.
S.P. acknowledges financial support from SERB-DST, Govt. of India via grant no.
SRG/2019/001419, and in the final stages of writing by
grant no. CRG/2021/003024.
The numerical results were obtained
using the computational facilities (Chandra cluster) of
the Department of Physics, IIT Bombay.
\end{acknowledgements}

\bibliographystyle{apsrev}
\bibliography{sign_prob_refs}


\end{document}